\documentclass[prb,twocolumn,floatfix]{revtex4-1} 
\usepackage{hyperref}
\usepackage{amsmath}  % needed for \tfrac, \bmatrix, etc.
\usepackage{amsfonts} % needed for bold Greek, Fraktur, and blackboard bold
\usepackage{graphicx} % needed for figures
\usepackage{float}
\usepackage{amssymb}
\usepackage{booktabs}
\usepackage{amsthm}
\usepackage{algorithmic}
\usepackage{mathtools}
\usepackage{physics}
\usepackage{comment}
\usepackage{textcomp}
\usepackage{xcolor}

\begin{document}

% Be sure to use the \title, \author, \affiliation, and \abstract macros
% to format your title page.  Don't use lower-level macros to  manually
% adjust the fonts and centering.

\title{Simple portable quantum key distribution for science outreach}
% In a long title you can use \\ to force a line break at a certain location.

\author{Pedro Neto Mendes}
%\email{} % optional
\affiliation{Departamento de Engenharia Electrotécnica e de Computadores, Instituto Superior Técnico, 1049-001, Lisbon, Portugal}
\affiliation{Instituto de Telecomunicações, \\ 1049-001, Lisbon, Portugal}

\author{Paulo André}
%\email{ajp@dickinson.edu}
\affiliation{Departamento de Engenharia Electrotécnica e de Computadores, Instituto Superior Técnico, 1049-001, Lisbon, Portugal}
\affiliation{Instituto de Telecomunicações, \\ 1049-001, Lisbon, Portugal}

\author{Emmanuel Zambrini Cruzeiro}
\email{emmanuel.zambrinicruzeiro@gmail.com}
\affiliation{Departamento de Engenharia Electrotécnica e de Computadores, Instituto Superior Técnico, 1049-001, Lisbon, Portugal}
\affiliation{Instituto de Telecomunicações, \\ 1049-001, Lisbon, Portugal}

%When submitting the manuscript for review, do not include the author's name or institution
%\author{Daniel V. Schroeder}
%\email{dschroeder@weber.edu} % optional
%\altaffiliation[permanent address: ]{101 Main Street, Anytown, USA} % optional second address
% If there were a second author at the same address, we would put another 
% \author{} statement here.  Don't combine multiple authors in a single
% \author statement.
%\affiliation{Department of Physics, Weber State University, Ogden, UT 84408-2508}
% Please provide a full mailing address here.

%\author{David P. Jackson}
%\email{ajp@dickinson.edu}
%\affiliation{Department of Physics, Dickinson College, Carlisle, PA 17013}

% See the REVTeX documentation for more examples of author and affiliation lists.

\date{\today}

\begin{abstract}
Quantum Key Distribution (QKD) has become an essential technology in the realm of secure communication, with applications ranging from secure data transmission to quantum networks. This paper presents a simple, compact, and cost-effective setup for undergraduate tutorial demonstrations of QKD. It relies on using weak coherent pulses, which can be readily produced using an attenuated laser. The system employs the simplified three-state BB84 protocol in free space, the states are encoded using linear polarization. Polarization encoding can be done passively or actively, depending on the budget available. Time multiplexing is implemented at the receiver to reduce the number of required detectors. Only two detectors are used to implement measurements on two bases, with a total of four outcomes. The result demonstrates the practicality of the system for free-space quantum communication, and its compact and portable nature makes it particularly suitable for pedagogical demonstrations. This work paves the way for engaging undergraduate students in the field of quantum communication through hands-on laboratory projects.
\end{abstract}
% AJP requires an abstract for all regular article submissions.
% Abstracts are optional for submissions to the "Notes and Discussions" section.

\maketitle % title page is now complete

\section{Introduction} % Section titles are automatically converted to all-caps.
% Section numbering is automatic.

Information security is a fundamental necessity of our interconnected world. It plays a pivotal role in numerous domains, including finance, military, industry, and even at the individual level. As the reliance on digital communication and information exchange continues to grow, traditional encryption methods, while effective in many scenarios, confront an approaching threat posed by the rapid advancement of quantum computing. These quantum computers have the potential to break through established encryption techniques, thereby raising concerns about the vulnerability of sensitive information.

This impending threat highlights the pressing need for a novel approach to secure communication, and Quantum Key Distribution (QKD) stands out as a pivotal solution. First proposed in 1984 \cite{BB84}, QKD offers the ability to exchange a secret key between two remote parties with proven security based only on the laws of physics. This key can later be used to communicate messages using protocols like the one-time-pad, offering a solution to protect secrecy in communication.

The current limitations in this technology revolve around scalability for longer distances and the practical challenges in implementation and, as such, the potential of the simplified three-state BB84 protocol,\cite{SimpleBB84} a novel protocol, to streamline and simplify setups becomes crucial. Understanding its capability to reduce complexity could be a significant step toward overcoming these limitations.

Moreover, to render QKD systems more practical, some assumptions can be relaxed, for example, the ideal single-photon source can be replaced by a source of attenuated laser pulses. At first glance, the use of these weak coherent pulses can make the system prone to photon number splitting (PNS) attacks. However, this security risk can be overcome with the use of decoy states,\cite{Decoy} which allow the users to determine more easily the presence of an eavesdropper. 

These practical choices allow for more compact, cheaper, and simpler setups that can be used in an undergraduate laboratory to explain how these protocols work while at the same time discussing fascinating questions that researchers have to deal with related to security issues and what really is quantum in a light pulse. Additionally, we also use time multiplexing to reduce the number of detectors required to implement the QKD protocol.

While educational articles exist on QKD implementation and single photon experiments in laboratories \cite{educational, single, double-slit, letter}, here we focus on the use of weak coherent pulses which offer a more straightforward and cost-effective setup, making quantum communication protocols accessible to undergraduate students. This work aims to bridge the gap between theory and practice, providing a valuable educational resource for understanding and implementing secure quantum communication.

\section{Theory}

\subsection{Basic quantum mechanics}

Quantum mechanics introduces fundamental principles governing the measurement of physical quantities within a system. When measuring a system, its physical quantity is represented by an operator, and the outcome corresponds to one of its eigenvalues. If the initial state of the system aligns with the measuring device's eigenstate, the measurement results in no change to the system. Conversely, if the initial state differs, the measurement "projects" the initial state onto one of the device's eigenstates, with the probability determined by the inner product of the initial and final states. This process leads to an irreversible alteration of the system, with the outcome inherently random.

A pure quantum bit, or qubit, is a normalized vector in $\mathbb{C}_2$. In the Dirac notation, such a vector is denoted $|\psi\rangle$ and is called a wavefunction. It represents the state of the physical system, in this case the qubit. In general, a qubit can be parametrized in spherical coordinates,
\begin{equation}
|\psi\rangle = \begin{pmatrix}
\cos\frac{\theta}{2}\\
e^{i\varphi}\sin\frac{\theta}{2}
\end{pmatrix}
\end{equation}
where $\theta$ is related to how close one is from the $|0\rangle$ (equiv. $|1\rangle)$ state, and $\varphi$ gives a phase which describes a rotation of the state around the z-axis.

The basis we implicitly chose, ${|0\rangle,|1\rangle}$ is called the Z basis. Other bases for qubits are the X (${|+\rangle,|-\rangle}$) and Y (${|R\rangle,|L\rangle}$) bases. When encoding a qubit in the linear polarization of light, $|0\rangle$ becomes $|H\rangle$, i.e. a horizontally polarized mode, $|1\rangle$ becomes $|V\rangle$, $|+\rangle$ becomes $|D\rangle$, and $|-\rangle$ becomes $|A\rangle$.

A measurement in quantum mechanics is represented by a set of operators ${E_k}$ ($C_2 \times C_2$ matrices). Note that there exist more general forms in our case we restrict to rank one projectors, which means they can be written in terms of one vector, i.e. $E_k=|k \rangle \langle k|$. The number of operators, or equivalently of indices $k$, represents the number of measurement outcomes. In this tutorial, all the measurements we consider have two outcomes. Finally, in quantum mechanics, one computes the probability of obtaining an outcome $k$ when measuring the state $|\psi\rangle$ with measurement $E_k$ using Born's rule, which reads $\text{Prob}(k) = \langle\psi|E_k|\psi\rangle$.

\subsection{One Time Pad and BB84}

The one-time pad (OTP) is an information-theoretically secure encryption technique, meaning that the encrypted message does not provide information about the original message. This communication method requires a novel, symmetric, and random key in each communication round and these keys should be larger than the message size.

The message $m$ is combined with the secret key $k$ through modular addition creating the cipher message $c$ ($c=m\oplus k$). The cipher message can be combined with the secret key to recover the original message ($m=c\oplus k$).

The difficulty of implementing this in practice comes from the creation of these keys between the communication parties. The BB84 protocol is a quantum key distribution scheme where the emitter (usually referred to as Alice) can share a secret key with the receiver (usually referred to as Bob). These parties are connected through a quantum channel and a classical channel, where they exchange quantum and classical signals. This channel can be attacked and eavesdropped by a third party, Eve.

Alice encodes her bit message on the quantum states of two possible bases, Z and X for example. A bit 0 is encoded on the state $|0\rangle$ or $|+\rangle$, for basis Z and X respectively. A bit 1 is encoded on the state $|1\rangle$ or $|-\rangle$, for basis Z and X respectively.

Bob chooses one of the bases and obtains a bit from the result. If Alice and Bob chose the same basis, they will share the same bit value. For example, if Alice sends bit 0 in the Z basis and Bob performs a measurement on that basis, the probability of Bob measuring bit 0 is given by

\begin{equation}
    \text{Prob}(0) = \langle 0 |E_0| 0 \rangle = \langle 0 | 0 \rangle \langle 0 | 0 \rangle = 1.
\end{equation}

Alternatively, if the measurement choice differs from Alice's basis, there exists an error probability. For example, if Alice sends bit 0 encoded on the Z basis and Bob measures on the X basis, the probability of Bob measuring bit 1 is

\begin{equation}
    \text{Prob}(-) = \langle 0 |E_-| 0 \rangle = \langle 0 | - \rangle \langle - | 0 \rangle = \frac{1}{2}.
\end{equation}

This means that Bob cannot recover the correct bit string when measuring on a different basis every round.

However, this measurement error, arising from a choice of different bases, can help us identify the presence of an eavesdropper (typically named Eve). If Eve intercepts and resends the state she measured from Alice, she will introduce errors. 

As Eve does not know which basis Alice and Bob chose, she will guess a measurement basis. If she chooses a basis different from Alice and Bob, she will prepare a different state than the original. Thus, Eve will introduce errors.

For example, if Alice sends bit 0 on the Z basis and Bob chooses the Z basis, the outcome will be bit 0 every time. If however, Eve performs a measurement on the X basis and prepares the corresponding state, Bob will have a 50\% chance of measuring the wrong outcome.

This is the idea of the BB84 protocol. It works as follows:
\begin{enumerate}
    \item \textbf{Quantum communication}: Alice creates a binary random key string and for each bit, randomly chooses on which basis to encode the information. She then sends through the quantum channel the respective quantum state. Bob makes a random basis choice, measures the state, and stores the output.
    
    \item \textbf{Sifting}: Alice and Bob share through the classic channel their respective basis choice. They then discard the bits for which the bases do not match and save the rest.
    
    \item \textbf{Post processing}: Alice and Bob estimate how much information was leaked to Eve through the errors on the sifted key, and perform error correction and privacy amplification. The protocol is aborted if Eve's presence is identified.
\end{enumerate}

\begin{table}[H]
\centering
\begin{tabular}{|c|c|c|c|c|c|c|c|}
\hline 
\textbf{Alice's bit} & 0 & 1 & 1 & 0 & 1 & 0 & 1 \\ \hline
\textbf{Alice's basis} & Z & Z & X & Z & X & X & Z \\ \hline
\textbf{Alice's state polarization} & $\mathrm{V}$ & $\mathrm{H}$ & $\mathrm{A}$ & $\mathrm{V}$ & $\mathrm{A}$ & $\mathrm{D}$ & $\mathrm{H}$\\ \hline \hline
\textbf{Eve's basis} & Z & X & Z & Z & X & Z & X\\ \hline
\textbf{Eve's state polarization} & $\mathrm{V}$ & $\mathrm{D}$ & $\mathrm{H}$ & $\mathrm{V}$ & $\mathrm{A}$ & $\mathrm{H}$ & $\mathrm{D}$ \\ \hline \hline
\textbf{Bob's basis} & Z & X & X & X & Z & X & Z \\ \hline
\textbf{Bob's state polarization} & $\mathrm{V}$ & $\mathrm{D}$ & $\mathrm{D}$ & $\mathrm{A}$ & $\mathrm{H}$ & $\mathrm{D}$ & $\mathrm{H}$ \\ \hline \hline
\textbf{Sifting} & \multicolumn{7}{c|}{} \\ \hline
\textbf{Shared secret key} & 0 &  & 0 &  &  & 0 & 1 \\ \hline
\textbf{Errors in key} & $\checkmark$ &  & $\times$ &  &  & $\checkmark$ & $\checkmark$ \\ \hline 
\end{tabular}
\caption{Shared secret key generated by the BB84 protocol in the presence of Eve.}
\label{tab:qkd}
\end{table}

Table \ref{tab:qkd} illustrates a few rounds of communication between Alice and Bob and how errors can appear in the shared key.

The error correction step is necessary to ensure that Alice and Bob share the same string of bits. This is accomplished using protocols like Cascade or LDPCs \cite{Error_corection}, which require an estimation of the quantum bit error rate (QBER) and additional information to be shared through the classical channel (e.g., the parity of blocks of the key string). This estimation of the QBER is done by sharing part of the key between Alice and Bob.
Finally, both parties run a privacy amplification protocol \cite{Renner} to ensure that even with the extra information Eve might have intercepted, she cannot reconstruct the secret key. This procedure reduces the size of the key.

The performance of the communication protocol can then be evaluated by some key parameters like the QBER, the secret key rate (SKR), and the transmission distance (given by how much loss the system can tolerate before the SKR becomes negligible). 

The QBER can be defined as the ratio of the number of erroneous bits to the total number of bits transmitted over a quantum channel. A lower QBER generally signifies a more secure and reliable QKD system, as it implies fewer errors in the transmitted key bits, increasing the potential for eavesdropping detection and the efficiency of the key generation process.

The SKR is the rate at which secure keys are generated and shared between two parties. It reflects the amount of usable key material produced per second after post-processing, ensuring secure communication. Higher secret key rates indicate more efficient and practical QKD systems.

The transmission distance is the distance at which the communication was performed and is related to the loss as for both free-space and fiber-based setups, the loss increases with the distance. The higher the loss an experimental setup can tolerate, the longer the transmission distance will be.

\subsection{Simplified BB84}

The simplified BB84 protocol is a variant of the original BB84 quantum key distribution protocol, designed to streamline the process while maintaining its security features. In the computational basis Z, the protocol runs exactly as the original BB84. However, in the monitoring basis X, the emitter prepares only $|D\rangle$, only three preparations are necessary. The protocol then runs similarly to the BB84:

\begin{enumerate}
    \item \textbf{Quantum communication}: Encoding is randomly done in Z and X bases. The emitter sends $|H\rangle$ and $|V\rangle$ uniformly in the Z basis, while in the X basis, it only emits $|D\rangle$. The receiver measures in X or Z bases with respective probabilities $p_X^B$ and $p_Z^B = 1 - p_X^B$. Basis choice and measurement outcome are recorded.
    
    \item \textbf{Sifting}: Both parties announce their chosen bases for each event. Z basis events are used to generate the raw key, while X basis events are utilized to estimate the presence of an eavesdropper. This step concludes after collecting a predetermined number of raw key bits.
    
    \item \textbf{Post processing}: An error correction algorithm is applied to the block of bits by both parties, during which some bits may be disclosed. Privacy amplification is applied to the block of bits to obtain a secret key.
\end{enumerate}

In the original BB84 protocol, the probabilities $p_X^B$ and $p_Z^B$ were fixed at 50$\%$, but various protocols and experimental setups can benefit from asymmetric measurement choices. For instance, in the simplified BB84 protocol, one basis monitors for eavesdropping while the other generates the key. Ideally, the monitoring basis measurements should be the minimal value that still allows for a complete monitor of the channel. 
%In practice, this is set to 10$\%$ due to the existence of readily available commercial devices available for this purpose.

For educational purposes focused on understanding the experimental implementation of secure communication through Quantum Key Distribution (QKD), a detailed description of error correction and privacy amplification was omitted. More information and the the security analysis can be found in .\cite{QBER}

\section{Experimental implementation}

The objective of this setup is to be practical, compact, and easy to construct. It enables a straightforward experiment for characterizing the setup, providing insights into the protocol's functionality without complete implementation. Additional equipment and effort can then be employed to implement the fully functioning protocol.

We selected the wavelength of 850 nm for the experiment due to its combination of efficient single-photon detection capabilities and the availability of commercial products operating at this wavelength.

This setup is divided into two parts as seen in the appendix \ref{appendixA}, two optical breadboards of 60x60 cm (Thorlabs, B6060A), the emitter, and the receiver. These are the breadboards available in the laboratory but smaller breadboards could be used instead. The emitter is capable of sending weak coherent pulses\cite{Coherent} in three different polarization states ($|\mathrm{H}\rangle,\ |\mathrm{V}\rangle,\ |\mathrm{D}\rangle$). The receiver in turn is capable of measuring on two different bases, $Z$ and $X$.

\begin{figure}[H]
    \centering
    \includegraphics[width=0.5\textwidth]{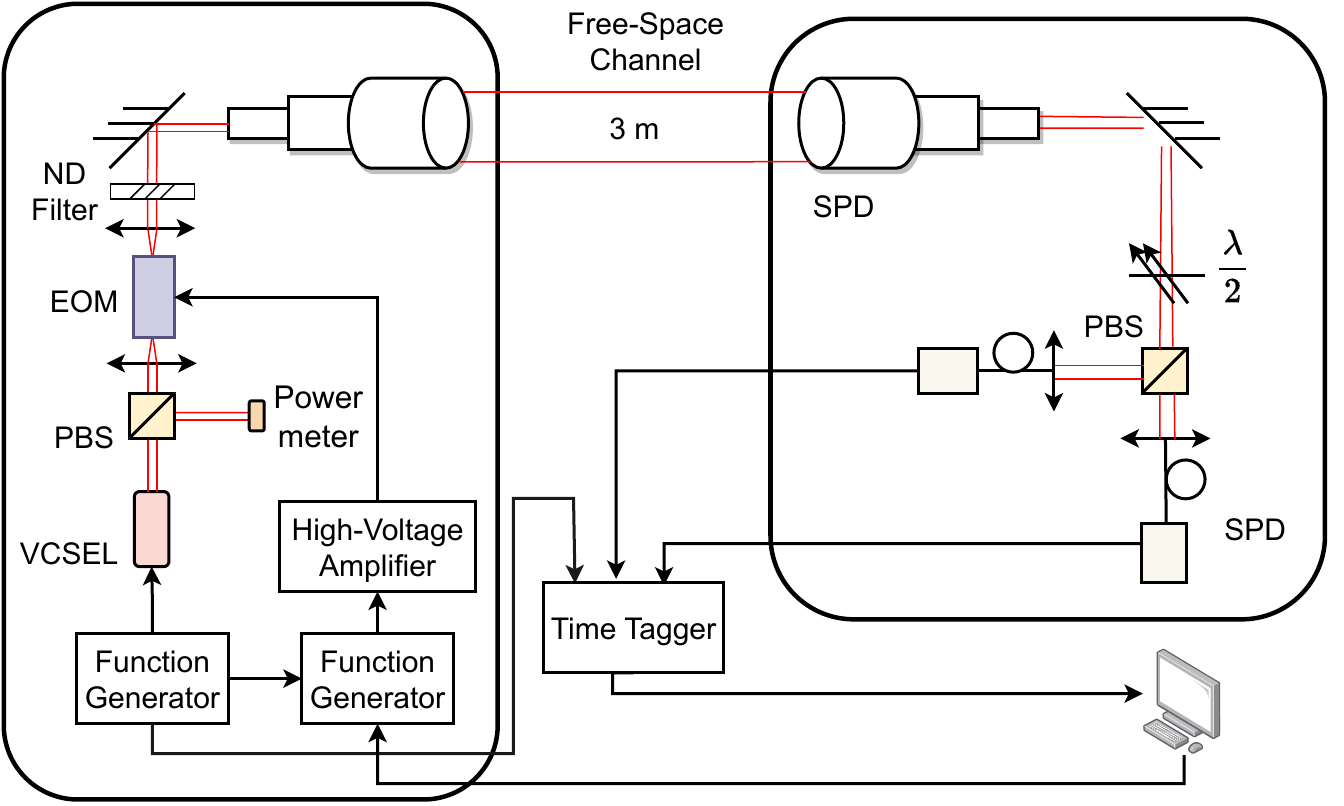}
    \caption{Simplified setup. Black arrows illustrate signal propagation between the devices. VCSEL: vertical-cavity surface-emitting laser. ND Filter: neutral density filters. SPD: single photon detector. PBS: polarizing beam splitter. EOM: electro-optical modulator. $\lambda$/2: half-wave plate.}
    \label{fig:simplesetup}
\end{figure}

\subsection{Emitter}

The goal of the emitter is to create three different quantum states. These will be weak coherent states with a given average number of photons per pulse and a chosen polarization. All states will have the same average number of photons with different polarization, $|\mathrm{H}\rangle$, $|\mathrm{V}\rangle$, and $|\mathrm{D}\rangle$.

As seen in Fig. \ref{fig:simplesetup}, the pulses are generated by directly modulating the vertical-cavity surface-emitting laser (VCSEL) (Roithner, VC850M2-MODULE) using a function generator (FC) (AIM-TTI, TG330). We opted for these components as they were readily available in the laboratory. Alternatively, a simple 850 nm laser and any function generator or driver capable of producing pulses could suffice. The pulse size depends on different experimental parameters and will be explained later.

After creating the pulses, the light is directed through a polarization beam splitter (PBS) (Thorlabs, PBS102), with the reflected power monitored by a power meter (Thorlabs, S120VC). This monitoring is used to estimate the number of photons emitted. The transmitted light is horizontally polarized and focused into an electro-optic modulator (EOM) (Thorlabs, EO-AM-NR-C1) with the input polarization adjusted to maximize the EOM's efficiency. This allows a fast (kHz) and active choice of polarization modulation.

While this setup requires a high-voltage amplifier (HVA) (Thorlabs, HVA200) to operate, an alternative option is the resonant EOM (Thorlabs, EO-PM-R-20-C1), which eliminates the need for a high-voltage amplifier.

The EOM creates distinct polarization states by applying varying voltages at the input. The specific voltage values required for the three different states necessary for the protocol depend on the alignment of the EOM crystal, which functions as a variable waveplate, effectively rotating the polarization of the light passing through it.

To achieve the desired polarization states within the available voltage range, a quarter-wave plate (QWP) (Thorlabs, WPQSM05-850) followed by a half-wave plate (HWP) (Thorlabs, WPHSM05-850) and another QWP can be employed to correct the output states of the EOM to match the targeted ones, simplifying the alignment. Additionally, due to our laser spot size being larger than the aperture of the EOM, we opted to utilize lenses (Thorlabs, LA1257-B) to minimize losses in this component.

These voltages are supplied by an arbitrary function generator (Agilent, 33250A), triggered by the first function generator, and the high-voltage amplifier. Alternatively, other function generators can be utilized if they allow the adjustment of the output voltage for each pulse and can be triggered externally. This flexibility enables encoding any bit sequence into photon polarization using the Z basis or the state in the X basis.

\begin{figure}[H]
    \centering
    \includegraphics[width=0.5\textwidth]{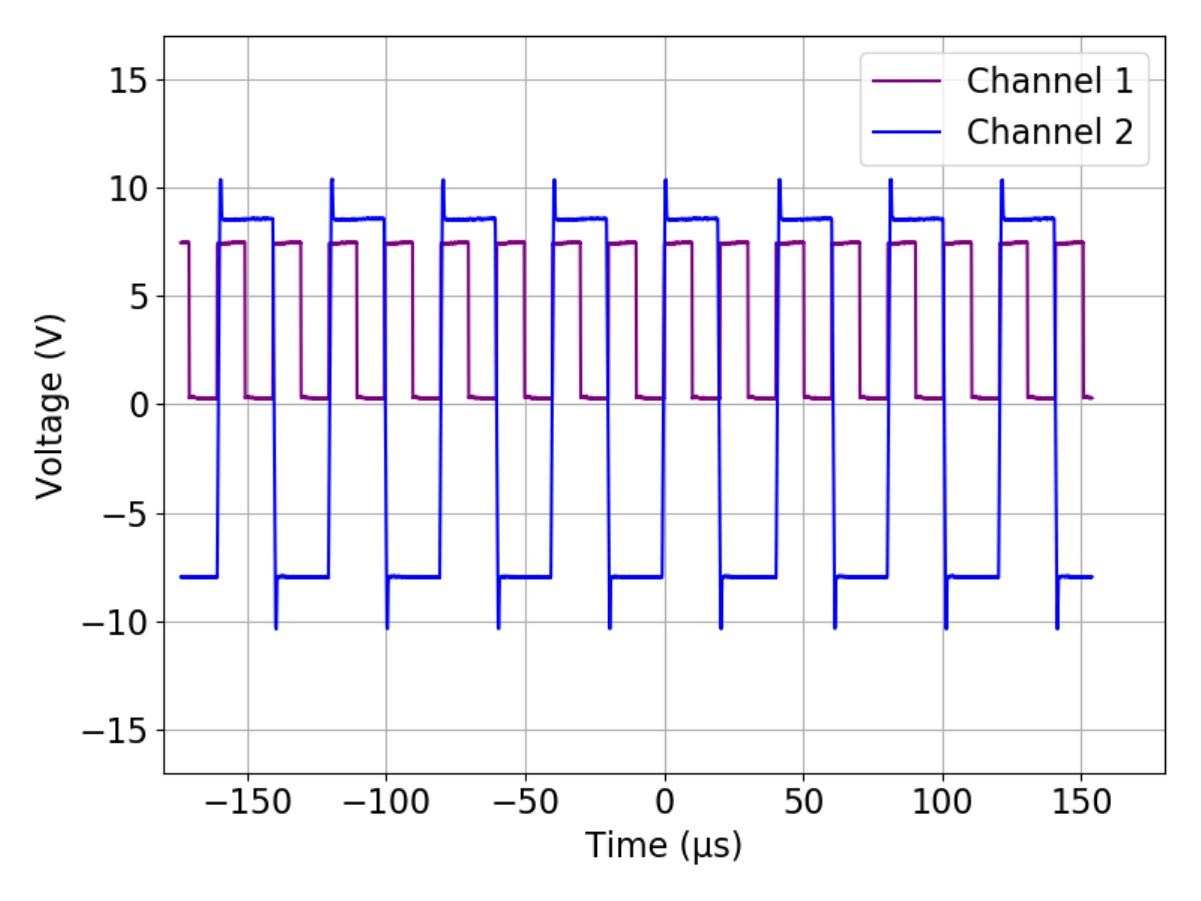}
    \caption{Electrical pulses seen in the oscilloscope. The purple pulse (channel 1) is used to modulate the VCSEL. The blue pulse (channel 2) is used to modulate the EOM to encode information in the pulses.}
    \label{fig:Osciloscope}
\end{figure}

Figure \ref{fig:Osciloscope} illustrates the electrical pulses employed for modulating the devices. The blue pulses represent the modulation of the EOM, responsible for polarization rotation. In this depiction, two distinct states are evident: the high voltage maintains horizontal polarization, while the low voltage pulse rotates it to a vertical orientation. Any slight distortion observed in the figure is attributable to impedance mismatch.

The purple pulses are utilized to modulate the VCSEL to switch the laser on and off. The synchronization shown in the figure demonstrates the simultaneous modulation of these two types of pulses: blue for polarization control and purple for laser activation. This synchronization enables independent polarization control for each light pulse. In this case, the pulse duration was 10 $\mu s$ and the repetition rate was 50 kHz.

Subsequently, the photons pass through neutral density filters (Thorlabs, NE50A) to adjust the average photon count per pulse to the desired value before being emitted into the free-space channel via a telescope (Thorlabs, GBE03-B). Filter selection depends on the input power and the specific parameters of the targeted weak coherent pulse and will be explained later.

\begin{figure}[H]
    \centering
    \includegraphics[width=0.50\textwidth]{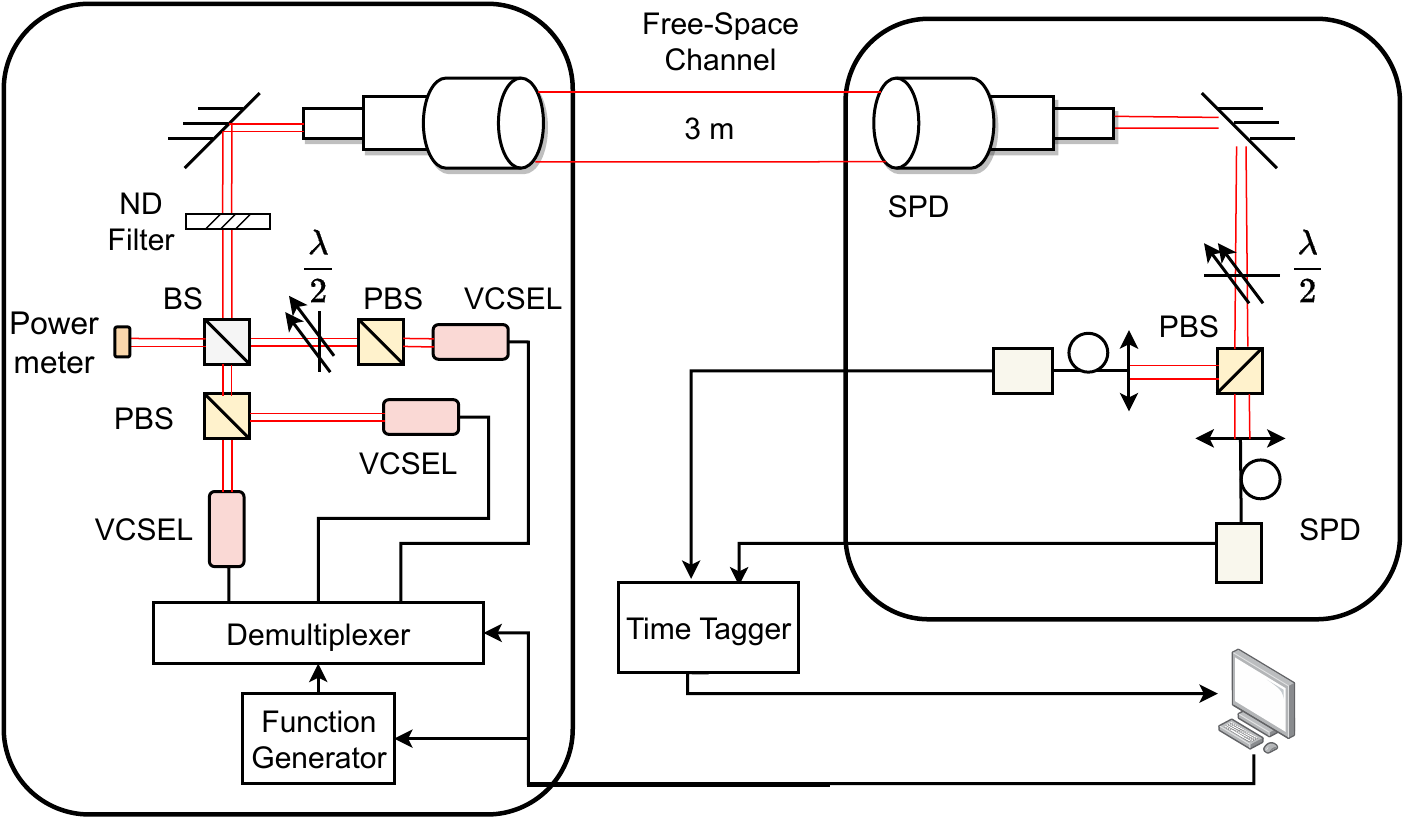}
    \caption{Simplified setup with a less costly emitter. Black arrows are electrical connections. VCSEL: vertical-cavity surface-emitting laser. ND filter: neutral density filters. SPD: single photon detector. PBS: polarizing beam splitter. $\lambda$/2: half-wave plate.}
    \label{fig:simplesetup_passive}
\end{figure}

Alternatively, it is also possible to implement a less costly approach, removing the EOM from the setup as seen in Fig. \ref{fig:simplesetup_passive}. Instead of one laser, three would be used. Only one laser would be turned on at any time and the path taken by the light would create the respective state. To create the $|\mathrm{H}\rangle$ and $\ |\mathrm{V}\rangle$ states, only a PBS would be required. To create the $|\mathrm{D}\rangle$ state, a PBS followed by a HWP can be used. This way, the state choice would be done by choosing which laser to turn on instead of by active modulation of the light. The different paths would then be recombined using a 50/50 beam splitter (BS)(Thorlabs, BS011) and the rest of the emitter would work as in the active approach.

This method brings security risks as the lasers can have manufacturing differences (spatial mode, frequency...) and an eavesdropper might take advantage of those but for this work, it can be a solution.

Given that both the emitter and receiver are situated within the same room, the free-space channel is defined as the distance between the telescopes at each end. While a well-collimated beam and a short free-space channel may render the use of telescopes optional, their inclusion enables easy scalability of the channel length by adjusting the distance between optical tables. The emitter can be seen in Fig. \ref{fig:Alice} in the appendix \ref{appendixA}.

\subsection{Receivers}

The objective of the receiver is to measure the incoming states on one of two basis. By calculating the QBER, we can detect the presence of potential eavesdroppers (Eve) and generate secret keys. For educational purposes focusing on the concept of key generation rather than actual key generation, this process can be simplified.

To accomplish this, as seen in Fig. \ref{fig:simplesetup}, the light is captured by a telescope and directed through a HWP, followed by a PBS. The angle in the HWP will define if the measure is on the Z basis or the X basis. Each output of the PBS is then coupled to a fiber and a single-photon detector (SPD) (Excelitas, SPCM-AQRH-10). This detector was pre-aligned for multi-mode fibers as it simplifies the alignment but can be requested for single-mode pre-alignment.

The arrival time of the photons is recorded using a time tagger (Swabian Instruments, Time Tagger 20), which offers precision beyond the requirement; a more affordable option such as an electronic circuit or a field programmable gate array microcontroller can be used as described in. \cite{Coincidence, Coincidence2} Additionally, the trigger pulse modulating the laser should be connected to the time tagger for post-processing of the measured data.

This setup only allows us to measure on one basis, there is no basis choice during the experiment. This will still allow us to calculate a QBER for each basis and simulate the presence of Eve.

\begin{figure}[H]
    \centering
    \includegraphics[width=0.50\textwidth]{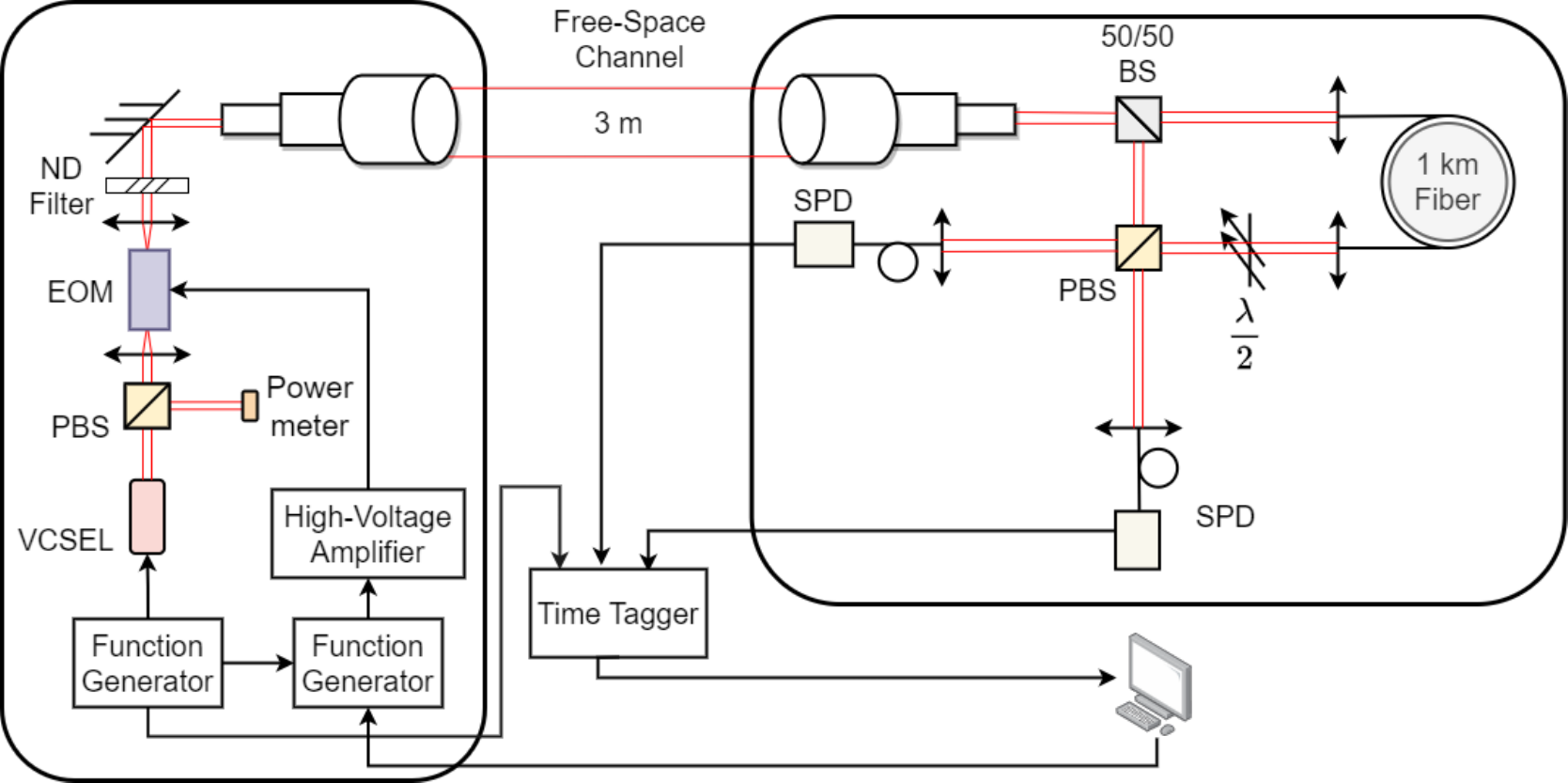}
    \caption{Complete receiver setup. Black arrows are electrical connections. VCSEL: vertical-cavity surface-emitting Laser. ND Filter: neutral density filters. SPD: single photon detector. BS: Beam Splitter. PBS: polarizing beam splitter. EOM: electro-optical modulator. $\lambda$/2: half-wave plate.}
    \label{fig:QKD_Setup}
\end{figure}

To improve the setup, we must allow for a basis choice in the receiver. To do this with only two detectors, we can use time multiplexing to separate in time the measurements on one basis from the measurements on the other basis. As seen in Fig. \ref{fig:QKD_Setup}, we start by adding a BS that will passively make the basis choice. Then each output path will correspond to a measure on one basis. The measurement in the Z basis does not require a HWP. The path of the X basis will need to rotate the measurement basis to X by applying a rotation with the HWP. 

In addition, one basis has to be delayed relative to the other. This allows us to distinguish which path the state took and on which basis it was measured. The beams are recombined and measured in the PBS and detected by the same SPDs as in the simplified setup.

The setup used can be seen in Fig. \ref{fig:Bob} in the appendix \ref{appendixA}.

\subsection{Parameter definition}

To implement the protocol, we need to define the pulse width of the coherent states, the pulse width of the signal modulating the EOM, the repetition rate, the average number of photons, and the time delay for the complete receiver.

The repetition rate will be limited by the high-voltage amplifier as it has a 1 MHz bandwidth. We chose a repetition rate of 50 kHz but a higher or lower value can be chosen.

For the simplified receiver, the pulse width can be chosen freely to be less than the repetition period. Our function generator can only generate square waves, resulting in a pulse width equal to half of the repetition period, 10 $\mu$s.

To determine the average number of photons in the weak coherent pulse, it is essential to thoroughly characterize the losses. By measuring the initial power and accounting for losses through various components, we can estimate the average number of photons detected by the receivers. 

In a secure implementation of the protocol, the average number of photons emitted from the emitter should be low. However, for educational purposes, we may increase this value to simplify the setup, as even with large values of loss, the number of photons arriving at the detector can replicate what would be observed in a secure QKD experiment (single photon regime). This may leave the communication open to attacks.

This means that the use of a non-optimal fiber (or other sources of loss as misalignment) would still allow for the experimental implementation as it would only require an increase in the number of photons sent, to compensate for the extra loss.

Defining $P$ as the initial laser power, $\Delta T$ as the pulse width, $E_{p}$ as the photon energy for this wavelength, and $\eta_{s}$ and $\eta_{f}$ as the loss of the setup, including here all the losses (components, coupling, channel), and the loss of the filters respectively, the average number of photons measured, $|\alpha|^2$ will be given by

\begin{equation}
    |\alpha|^2=\frac{P\Delta T}{E_{p}}10^{-\frac{\eta_{s}+\eta_{f}}{10}}.
    \label{eq:number of photons}
\end{equation}

From Eq.(\ref{eq:number of photons}), the necessary filter loss in dB can be determined. The parameters used in this work were: $\Delta T = 10 \mu s$, $P = 92 \mu W$, $\lambda = 850$ nm, $\eta_{s} = 12.2$ dB (including all the channel and receiver losses), and $\eta_{f} = 80$ dB. These values result in a predicted $|\alpha|_{Alice}^2 = 39.3$ (weak coherent state emitted by Alice) and $|\alpha|_{Bob}^2 = 2.35$ (weak coherent state arriving at the detectors), while the measured value was $2.30 \pm 0.05$. The measured probabilities of different photon counts were as follows: "0" with a probability of 0.11, "1" with 0.23, "2" with 0.25, and "3" with 0.19.

To implement the complete receiver, time multiplexing for detections is essential. Measurements in one basis must be temporally separated from those in the orthogonal basis to allow for efficient processing. Ideally, this requires a delay larger than the optical pulse width to prevent overlap between the detection windows of delayed and not delayed photons, facilitating the differentiation of measurement bases based on photon arrival time.

To achieve this with only two detectors, measurements on the X basis are deliberately delayed. Despite the lack of refractive index information for the fiber used, typical values generally fall within the range of [1.45, 1.48]. Utilizing an available 1590-meter multi-mode optical fiber, although not ideal, yields an expected delay of approximately [7.685, 7.844] $\mu$s (around 5 ns per meter of fiber). As the number of photons can be adapted in this demonstration, the loss in the fiber is not a concern.

To verify this delay and assess time-multiplexed detection, a low-photon-count pulse with diagonal polarization is emitted from the emitter. For experimental validation, the BS in the setup (Fig \ref{fig:QKD_Setup}) is replaced with a PBS to ensure minimal overlap between delayed and not delayed beams. The horizontal component of the polarized beam traverses a second PBS, reaching only detector 1 after passing through a filter to simulate losses in the alternate path, due to the fiber used. Conversely, the vertical component is directed into the optical fiber for temporal delay. Subsequently, a HWP rotates the polarization at the fiber output to horizontal, guaranteeing that photons exclusively reach detector 2 via a second PBS.

Consequently, clicks registered in detector 1 originate from the non-delayed signal, while those in detector 2 result from the delayed signal.

\begin{figure}[H]
    \centering
    \includegraphics[width=0.5\textwidth]{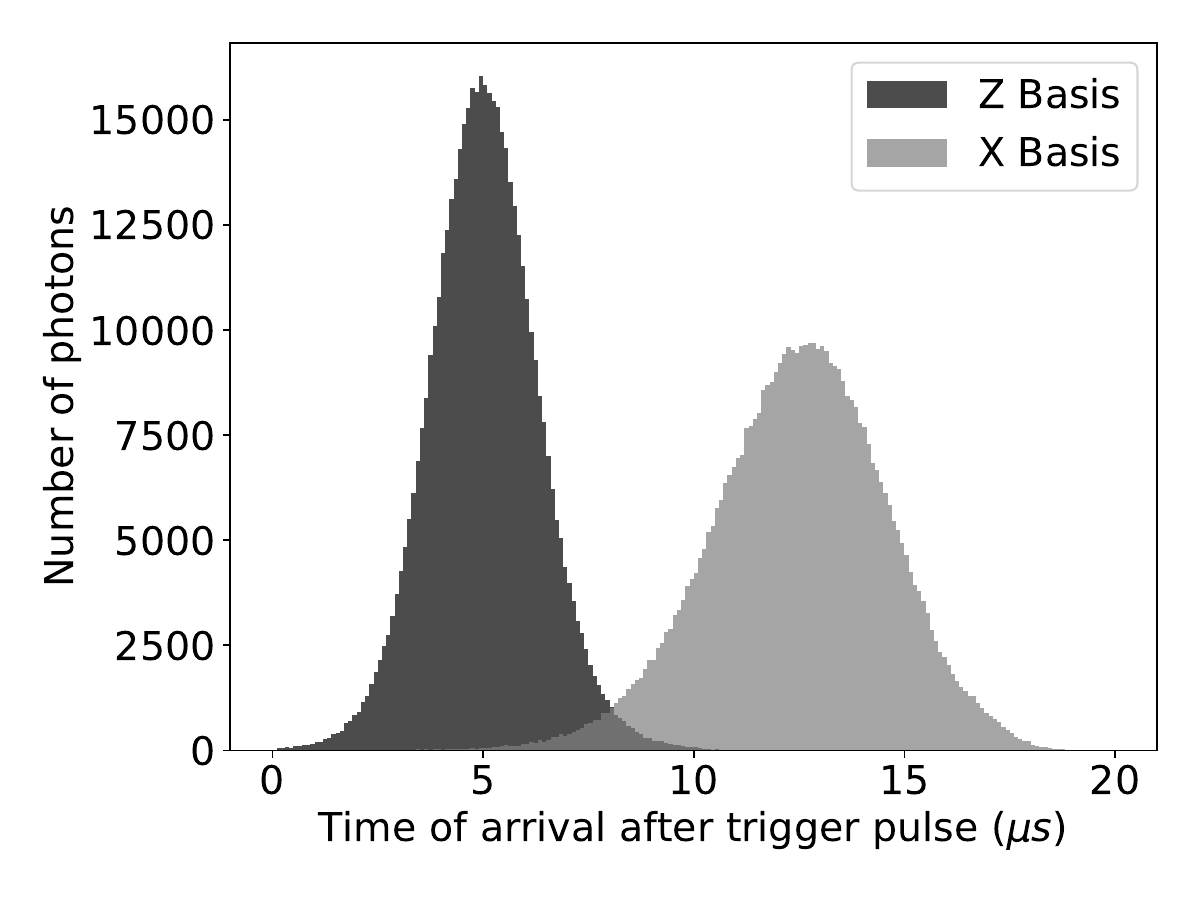}
    \caption{Histogram of the time of arrival of the photons after the trigger signal. The total number of photons is 491004 for detector 1, Z basis (in black), and 491071 for detector 2, X basis (in gray).}
    \label{fig:delay}
\end{figure}

In Fig. \ref{fig:delay}, the distribution of photon arrival times relative to the trigger is depicted. Two distinct normal distributions, centered around different values, are evident. The first, approximately $5.0 \pm 1.1$ $\mu$s, corresponds to the beam directly transmitted to the detector. The second, approximately $12.3 \pm 2.2$ $\mu$s, represents the delayed beam. Bars in gray are clicks in the second detector, while those in black are clicks in the first detector. The average delay between the two distributions is calculated to be $7.3 \pm 3.3$ $\mu$s.

The first normal distribution matches the expected outcome. As the pulse width is $10.0$ $ \mu$s and the events follow a poisson distribution, there is a higher likelihood of registering an event at the center of the pulse width and zero probability of observing outside the pulse width, creating the observed normal distribution. The second distribution shows a broader width, likely due to dispersion in the fiber. 

The observed overlap between the distributions can be attributed to selecting a delay shorter than the pulse width and to the broadening in the fiber, as now the normal distribution will have an event probability outside the original pulse width range. Despite this overlap, it remains minimal, allowing for clear differentiation between delayed and non-delayed signals, thereby facilitating the measurement of states in different bases.

\section{Results and analysis}

With the previously described setup, communication can be implemented. For simplicity, pre-determined messages can be encoded into the quantum states. In a secure implementation, an encoding scheme and random basis selection and message generation would be employed. The most straightforward message consists of a repeating string of alternating 0 s and 1 s throughout the communication ("0101010101..."). For the evaluation of the setup, the QBER on each basis must be measured. To simplify this, the message can be encoded solely on the Z basis and then, the experiment is repeated on the X basis. For a more complete demonstration, the complete receiver can be used and the initial message has to be encoded on both bases, basis selection is necessary.

\subsection{Simplified receiver}

The message is transmitted using the quantum states and detected using the HWP set to 0º for the Z basis and 22.5º for the X basis. Since the states are orthogonal, only one detector should register a click for each state sent. 

The single photon detectors create an electric pulse when an event is registered. This pulse will generate an event in the connected channel of the time tagger. This way, all the information is registered in the output file of the time tagger which includes the timestamp, the channel identification, and an error check parameter to access the correct use of the time tagger (missed events).

For the Z basis measurement, if the detector connected to the horizontal output of the PBS registers a click, the state sent was $|\mathrm{H}\rangle$, while a click in the other detector indicates the state was $|\mathrm{V}\rangle$. For educational purposes, it may be advantageous to define a higher average number of photons to ensure consistent detector clicks, simplifying the subsequent analysis. For the X basis, the same logic applies.

\begin{figure}[H]
    \centering
    \includegraphics[width=0.5\textwidth]{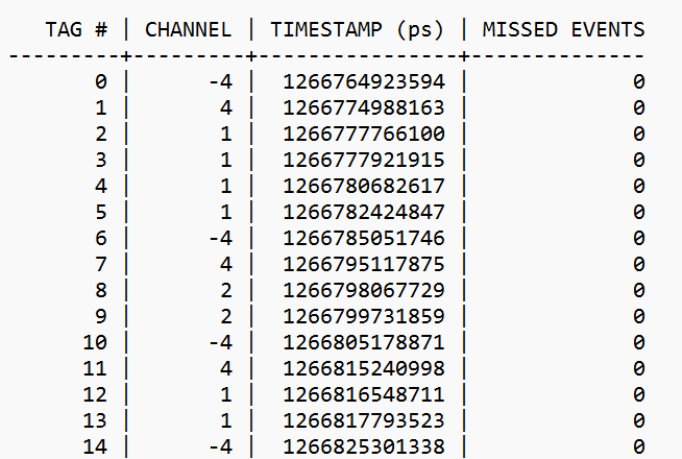}
    \caption{Example of events registered by the time tagger.}
    \label{fig:time_tagger}
\end{figure}

In Fig. \ref{fig:time_tagger}, an illustration of events captured by the time tagger is provided. Channel 4 monitors the modulation of the VCSEL, where event 4 represents a rising edge, indicating the onset of a pulse, while event -4 means a falling edge, denoting the end of the pulse duration. Therefore, the pulse is considered active between a 4 and a -4. Subsequently, the SPDs corresponding to the horizontal and vertical paths of the PBS are connected to channels 1 and 2, respectively.

Assuming that bit 0 is transmitted by the horizontal polarization and bit 1 is transmitted by the vertical state, by analyzing which detector clicked more, a guess of the bit sent can be performed. If there are more events on channel 1, then the state sent was $|\mathrm{H}\rangle$, and therefore, the bit sent by that pulse was 0. The same logic applies to channel 2. By doing this to all pulses, a recovered message can be found. 

Using this logic, from Fig. \ref{fig:time_tagger}, the string "010" is recovered, which matches the bits sent for this fraction of the message. By collecting data for 10 seconds, a message of 500 kbit is sent by Alice and recovered by Bob. By comparing this message to the original, counting the number of different bits, and dividing by the size of the message, the QBER can be found. We measure the QBER to be 2.5\% with an uncertainty of 0.4\% on the Z basis, while on the X basis, it is 2.11\% with an uncertainty of 0.14\%. These error rates are acceptable but could be reduced by addressing the impact of dark counts.

The electrical pulse is relatively wide, limited by the function generator available, and consequently, the detection window, the period during which signals are observed, is also broad. As a result, there's a higher likelihood of dark counts affecting the measured results. To improve this, one could lower the dark counts or reduce the pulse width and the detection window.

While this implementation did not focus on optimizing the SKR, it is still an important figure of merit. As we did not implement an error correction code and privacy amplification protocol we estimated the secret key size using the upper bound provided in.\cite{SimpleBB84}

As we used a weak coherent state with a high number of photons, the corresponding secret key rate is 0. The communication is open to attacks as Eve can capture a few photons each round and recover the entire message. For a secure implementation, the number of photons sent by Alice has to be reduced, and by using a python package created to estimate the SKR of the simplified BB84 for space-based communication, provided in\footnote{The code is available on the Github page \url{https://github.com/QuLab-IT/QuantSatSimulator.git}.}, we arrived at the optimal secret key rate of 75 bps for an average photon number of 0.87 and parameters reported in the appendix \ref{appendixB}. This value can be easily improved by increasing the source rate or the communication time window.

 Comparing these results to the ones reported in \cite{SimpleBB84}, it can be seen that this setup provides a much lower SKR. For the loss value measured in this experiment and a 625 MHz source, they achieve a QBER of around 3$\%$ and a SKR of around 8 kbps where the result is mainly limited by the detectors used as for low loss values their SKR saturates. 
 
 Recent works have shown a SKR of 115.8 Mbps over 10-km standard fiber for a source of 2.5 GHz, proving that high-rate quantum key distribution is possible \cite{high_rate}. There are also commercially available products like the Clavis XGR QKD platform from ID Quantique operating with a 1 Gbit source and generating secret keys at around 12 dB of loss at a rate of 400 kbps.\footnote{Clavis XGR QKD \url{https://www.idquantique.com/quantum-safe-security/products/clavis-xgr-qkd-platform/}.}

\subsection{Complete receiver}

If the complete receiver is used, the message sent has to have bits encoded in both bases, closer to what is used in a secure QKD setup. The procedure is similar to the simplified receiver, we obtain the data from the time tagger and must recover the message. In this receiver, we have to recover which basis was chosen and the message sent. To recover the basis choice, we look at the number of events delayed and not delayed. This means that the timestamp data shown in Fig. \ref{fig:time_tagger} has to be used. If there are more delayed events, events 10 $\mu s$ after an event in channel 4, the basis chosen is X, and if there are more photons without delay, events between the timestamp of channel 4 and 10 $\mu s$ after, the basis chosen is Z.

After, we can recover the message as done previously, looking into which detector clicked more. The basis choice can then be compared to Alice's choice and only the events with a basis match are kept. Finally, the QBER is calculated in the same way, by comparing both messages.

This is a simplified way to estimate the QBER where the whole message is shared between Alice and Bob, not just a fraction. This is followed again by the post-processing to generate the secret key.

\subsection{Simulating Eve}

To simulate the presence of Eve one can create a very simple experiment. If we periodically block the free space channel during the communication phase, we can simulate an attack where Eve removes photons from the channel. This will stop all detection, leading to an increase in the QBER. For example, if for no detection we assume bit 0 was sent, the message recovered while Eve is blocking the channel would be an infinite string of 0s.

\begin{figure}[H]
    \centering
    \includegraphics[width=0.5\textwidth]{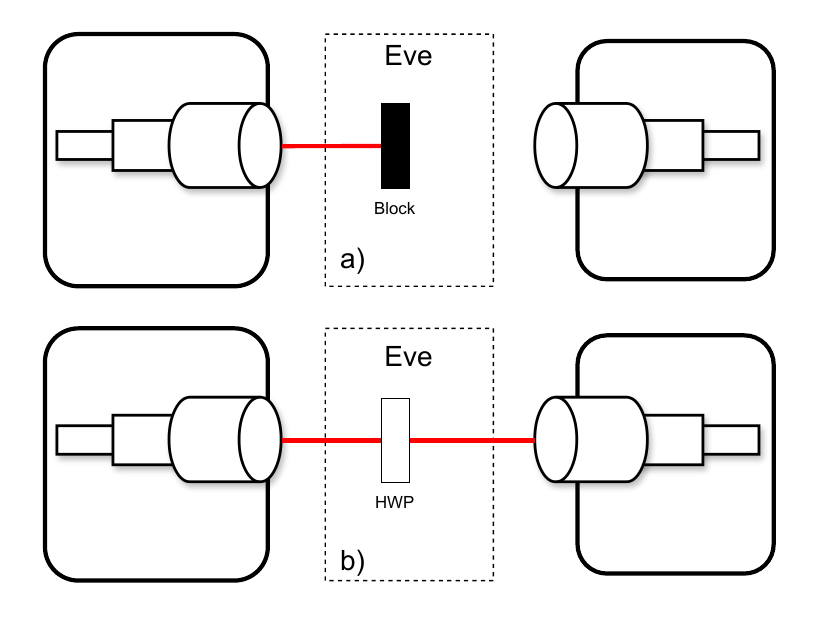}
    \caption{Design of Eve simulation experiments. a) The free space channel is blocked. b) A HWP is used to rotate the basis.}
    \label{fig:eve}
\end{figure}

Another possible approach is to use a HWP to rotate the state polarization. If we use an HWP at 22.5º, we can simulate an Eve that always chooses the wrong measurement basis, increasing the QBER to around 50\% as the message recovered would be random (the measurement is made on X basis while the state was encoded on the Z basis for example). This gives intuition on an intercept and resend attack where the basis choice is always wrong. The change in the experimental setup can be seen in Fig. \ref{fig:eve}.

These simple experiments show that Eve's presence in the channel leads to an increase in the QBER that can be detected aborting the communication and discarding the key generated as it was not secure.

\section{Conclusion}

In summary, this work has demonstrated a practical, compact, and cost-effective QKD setup suitable for undergraduate tutorial demonstrations. By using weak coherent pulses, the simplified three-state BB84 protocol, and time multiplexing, the complexity of the setup can be reduced and QKD demonstrations can be easily implemented, making the setup more accessible and manageable within educational environments. 

\section*{Author Declarations}

The authors have no conflicts to disclose.

\section{Acknowledgments}

E.Z.C. acknowledges funding by FCT/MCTES - Fundação para a Ciência e a Tecnologia (Portugal) - through national funds and when applicable co-funding by EU funds under the project UIDB/50008/2020 and 2021.03707.CEECIND/CP1653/CT0002.
The authors thank the support from the European Commission (EC) through project PTQCI (DIGITAL-2021-QCI-01).

\clearpage

\appendix

\section{Equipment Price}
\label{appendixA}

Here we compile the prices and quantities of the essential components to implement the emitter and complete receiver. It does not include all the equipment necessary, i.e., fibers, mechanical supports, and more.

\begin{table}[H]
    \centering
    \scalebox{0.85}{
    \begin{tabular}{ccccc} % Adjust the alignment and number of columns as needed
        \toprule
        \hline
        \hline
        \textbf{Designation} &  \textbf{Distributor} &  \textbf{Model} &  \textbf{Qt} & \textbf{Unit Price (\$)} \\
        \hline
        \midrule
        VCSEL & Roithner & VC850M2 & 1 & 75 \\
        FC & AIM-TTI & TG330 & 1 & 500 \\
        PBS & Thorlabs & PBS102 & 2 & 222 \\
        BS & Thorlabs & BS011 & 1 & 205 \\
        EOM & Thorlabs & EOAM-NR-C1 & 1 & 2999 \\
        HVA & Thorlabs & HVA200 & 1 & 3000 \\
        Arbitrary FC & Agilent & 33250A & 1 & 2975 \\
        ND Filter & Thorlabs & NE50A & 3 & 60 \\
        Telescope & Thorlabs & GBE03-B & 2 & 576 \\
        HWP & Thorlabs & WPHSM05-850 & 1 & 507 \\
        SPD & Excelitas & SPCM-AQRH-10 & 2 & 3110 \\
        Arbitrary FC & Thorlabs & 33250A & 1 & 2975 \\
        Time Tagger & Swabian & Time Tagger 20 & 1 & 10000 \\
        \hline
        \hline
        \bottomrule
    \end{tabular}}
    \caption{Equipment Prices in February 2024.}
    \label{tab:equipment_prices}
\end{table}

Some of the equipment in Table \ref{tab:equipment_prices} can be replaced by homemade solutions or cheaper commercial products as previously mentioned.

\begin{figure}[H]
    \centering
    \includegraphics[width=0.4\textwidth]{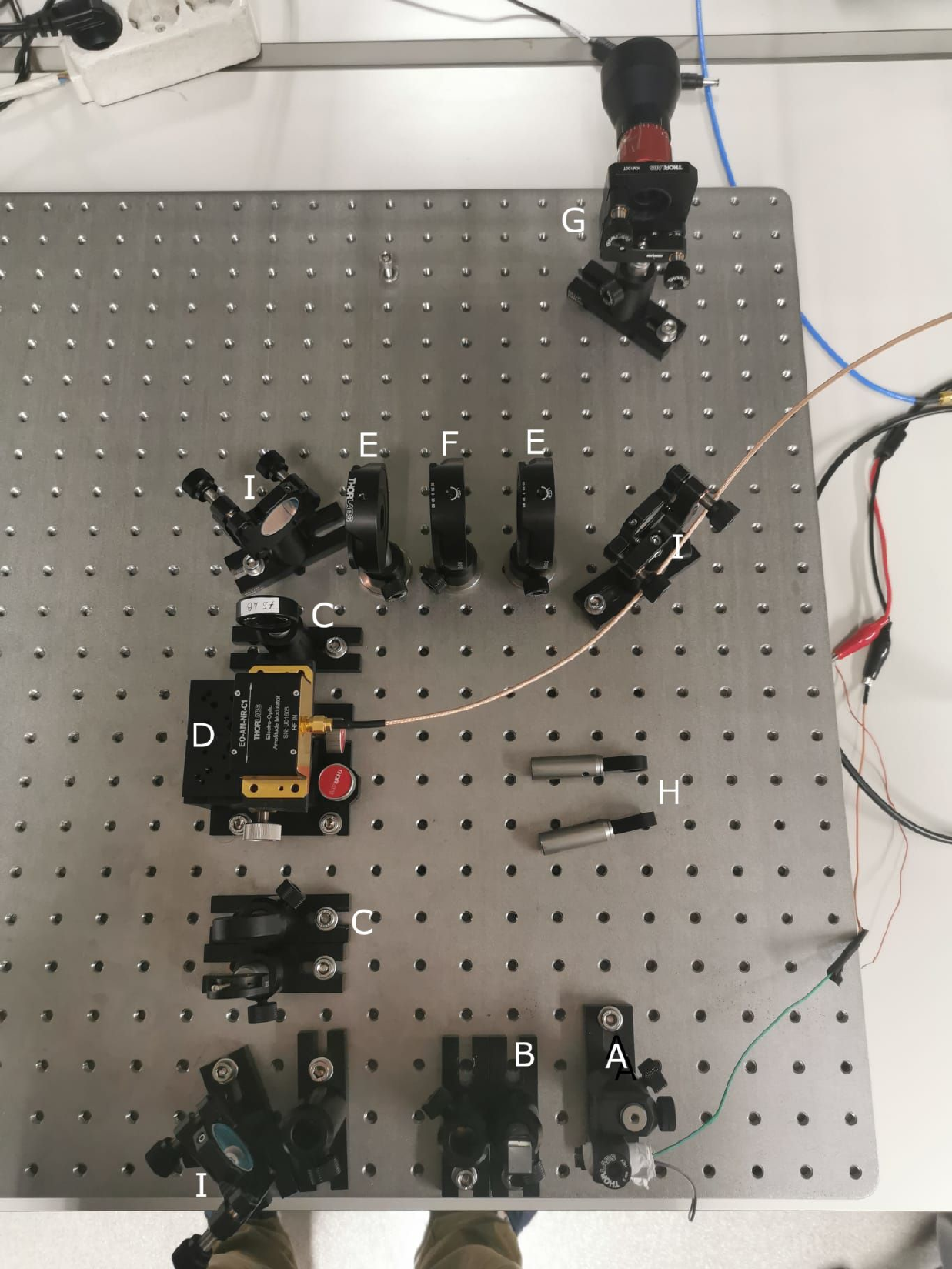}
    \caption{Photo from the emitter setup. A - VCSEL. B - PBS. C - Lens. D - EOM. E - QWP. F - HWP. G - Telescope. H - ND Filters I - Mirror.}
    \label{fig:Alice}
\end{figure}

\begin{figure}[H]
    \centering
    \includegraphics[width=0.4\textwidth]{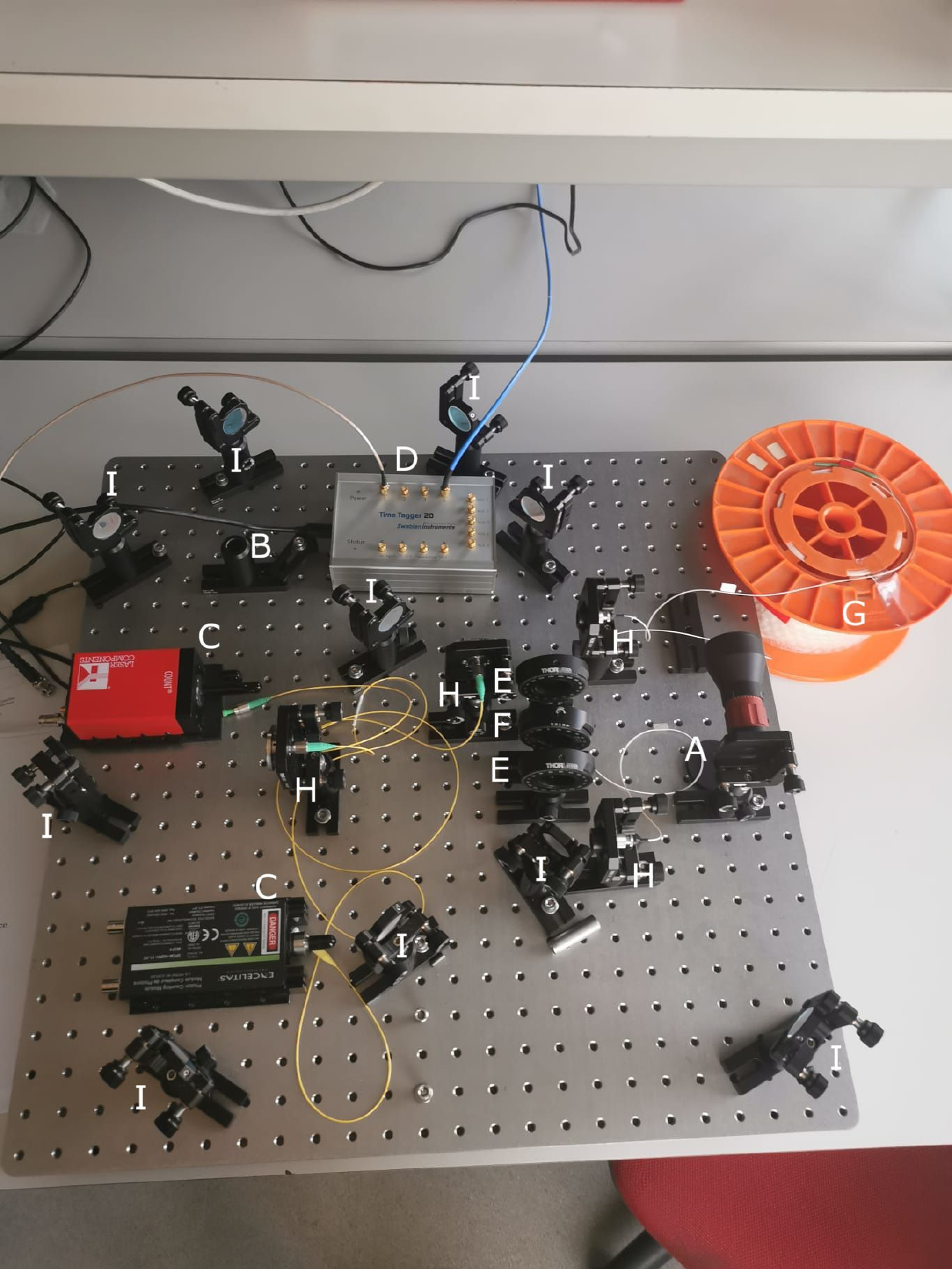}
    \caption{Photos from the complete receiver setup. A - Telescope. B - PBS. C - SPD. D - Time Tagger. E - QWP. F - HWP. G - Delay Fiber. H - Fiber couplers I - Mirror.}
    \label{fig:Bob}
\end{figure}

In Fig. \ref{fig:Alice} and Fig. \ref{fig:Bob} the experimental setups used can be seen.

\section{Parameters for SKR Optimization}
\label{appendixB}

The parameters used for the optimization of the SKR for this setup is shown in Table \ref{tab:optparam}.

\begin{table}[H]
\centering
\begin{tabular}{ |c|c| }
\hline
Parameter & Value\\
\hline
Signal intensity & 0.79 \\
\hline
Probability of sending signal & 0.75 \\
\hline
Decoy intensity & 0.02 \\
\hline
Probability Alice sends an $Z$ basis signal & 0.5 \\
\hline
Probability Bob measures an $Z$ basis signal & 0.5 \\
\hline
Loss & 12 dB \\
\hline
Communication time & 10 s \\
\hline
Correctness parameter ($ \epsilon_C $) & $10^{-15}$ \\
\hline
Secrecy parameter ($\epsilon_S$) & $10^{-9}$ \\
\hline
Intrinsic Quantum Bit Error Rate & $10^{-3}$ \\
\hline
Extraneous count probability & $10^{-3}$ \\
\hline
After pulse probability  & $10^{-3}$ \\
\hline
Source repetition rate & 100 kHz \\
\hline
\end{tabular}
\caption{Parameters used for SKR optimization}
\label{tab:optparam}
\end{table}


\begin{thebibliography}{2}%
\makeatletter
\providecommand \@ifxundefined [1]{%
 \@ifx{#1\undefined}
}%
\providecommand \@ifnum [1]{%
 \ifnum #1\expandafter \@firstoftwo
 \else \expandafter \@secondoftwo
 \fi
}%
\providecommand \@ifx [1]{%
 \ifx #1\expandafter \@firstoftwo
 \else \expandafter \@secondoftwo
 \fi
}%
\providecommand \natexlab [1]{#1}%
\providecommand \enquote  [1]{``#1''}%
\providecommand \bibnamefont  [1]{#1}%
\providecommand \bibfnamefont [1]{#1}%
\providecommand \citenamefont [1]{#1}%
\providecommand \href@noop [0]{\@secondoftwo}%
\providecommand \href [0]{\begingroup \@sanitize@url \@href}%
\providecommand \@href[1]{\@@startlink{#1}\@@href}%
\providecommand \@@href[1]{\endgroup#1\@@endlink}%
\providecommand \@sanitize@url [0]{\catcode `\\12\catcode `\$12\catcode `\&12\catcode `\#12\catcode `\^12\catcode `\_12\catcode `\%12\relax}%
\providecommand \@@startlink[1]{}%
\providecommand \@@endlink[0]{}%
\providecommand \url  [0]{\begingroup\@sanitize@url \@url }%
\providecommand \@url [1]{\endgroup\@href {#1}{\urlprefix }}%
\providecommand \urlprefix  [0]{URL }%
\providecommand \Eprint [0]{\href }%
\providecommand \doibase [0]{http://dx.doi.org/}%
\providecommand \selectlanguage [0]{\@gobble}%
\providecommand \bibinfo  [0]{\@secondoftwo}%
\providecommand \bibfield  [0]{\@secondoftwo}%
\providecommand \translation [1]{[#1]}%
\providecommand \BibitemOpen [0]{}%
\providecommand \bibitemStop [0]{}%
\providecommand \bibitemNoStop [0]{.\EOS\space}%
\providecommand \EOS [0]{\spacefactor3000\relax}%
\providecommand \BibitemShut  [1]{\csname bibitem#1\endcsname}%
\let\auto@bib@innerbib\@empty
%</preamble>
\bibitem [{Note1()}]{Note1}%
  \BibitemOpen
  \bibinfo {note} {The code is available on the Github page \protect \url {https://github.com/QuLab-IT/QuantSatSimulator.git}.}\BibitemShut {Stop}%
\bibitem [{Note2()}]{Note2}%
  \BibitemOpen
  \bibinfo {note} {Clavis XGR QKD \protect \url {https://www.idquantique.com/quantum-safe-security/products/clavis-xgr-qkd-platform/}.}\BibitemShut {Stop}%
\end{thebibliography}%


\begin{thebibliography}{3}


\bibitem{BB84} Bennett C. H. and Brassard G., ``Quantum cryptography: Public key distribution and coin tossing.'' Theoretical computer science \textbf{560}, 7--11 (2014).

\bibitem{SimpleBB84} Grünenfelder F., Boaron A., Rusca D., Martin A. and Zbinden H., ``Simple and high-speed polarization-based QKD.'' Applied Physics Letters \textbf{112}(5) (2018).

\bibitem{Decoy} Ma X., Qi B., Zhao Y. and Lo H. K. ``Practical decoy state for quantum key distribution.'' Physical Review A, \textbf{72}(1), 012326 (2005).

\bibitem{educational} Bista A., Sharma B. and Galvez E. J., ``A demonstration of quantum key distribution with entangled photons for the undergraduate laboratory.'' American Journal of Physics \textbf{89}(1), 111--120 (2021).

\bibitem{single} Pearson B. J. and Jackson D. P., ``A hands-on introduction to single photons and quantum mechanics for undergraduates.'' American Journal of Physics \textbf{78}(5), 471-484 (2010).

\bibitem{double-slit} Rueckner W. and Peidle J. ``Young's double-slit experiment with single photons and quantum eraser.'' American Journal of Physics \textbf{81}(12), 951-958 (2013).

\bibitem{letter} Galvez E. J. ``Resource letter SPE-1: Single-photon experiments in the undergraduate laboratory.'' American Journal of Physics \textbf{82}(11), 1018-1028 (2014).

\bibitem{Error_corection} Johnson J. S., Grimaila M. R., Humphries J. W. and Baumgartner G. B. ``An analysis of error reconciliation protocols used in quantum key distribution systems.'' The Journal of Defense Modeling and Simulation, \textbf{12}(3), 217-227 (2015).

\bibitem{Renner} Renner R. ``Security of quantum key distribution.'' International Journal of Quantum Information, \textbf{6}(01), 1-127 (2008).

\bibitem{QBER} Boaron A., Korzh B., Houlmann R., Boso G., Lim C. C. W., Martin A. and Zbinden H., ``Detector-device-independent quantum key distribution: Security analysis and fast implementation.'' Journal of Applied Physics, \textbf{120}(6) (2016).

\bibitem{Coherent} Jean-Pierre Gazeau, \textit{Coherent States in Quantum Physics}, 1st edition (Wiley, 2009).

\bibitem{Coincidence} Branning D., Bhandari S. and Beck M. ``Low-cost coincidence-counting electronics for undergraduate quantum optics.'' American Journal of Physics \textbf{77}(7), 667-670 (2009).

\bibitem{Coincidence2} Galvez E. J., Holbrow C. H., Pysher M. J., Martin J. W., Courtemanche N., Heilig L. and Spencer J. ``Interference with correlated photons: Five quantum mechanics experiments for undergraduates.'' American Journal of Physics \textbf{73}(2), 127-140 (2005).

\bibitem{high_rate} Li W., Zhang L., Tan H., Lu Y., Liao S. K., Huang J., ... and Pan J. W. ``High-rate quantum key distribution exceeding 110 Mb s–1.'' Nature Photonics, \textbf{17}(5), 416-421 (2023).


\end{thebibliography}
\end{document}